# Stimulated emission depletion microscopy with array detection and photon reassignment

**Wensheng Wang,[1] Zhimin Zhang,[1] Shaocong Liu,[1] Yuchen Chen,[1] Liang Xu,[1] Cuifang Kuang[1,2,*] Xu Liu[1,2]**

[1]State Key Laboratory of Modern Optical Instrumentation, Department of Optical Engineering, Zhejiang University, Hangzhou 310027, China
[2]China Collaborative Innovation Center of Extreme Optics, Shanxi University, Taiyuan, 030006, China
*Corresponding author: cfkuang@zju.edu.cn

**We propose a novel stimulated emission depletion (STED) microscopy based on array detection and photon reassignment. By replacing the single-point detector in traditional STED with a detector array and utilizing the photon reassignment method to recombine the images acquired by each detector, the final photon reassignment STED (prSTED) image could be obtained. We analyze the principle and imaging characteristics of prSTED, and the results indicate that, compared with traditional STED, prSTED can improve the signal-to-noise ratio (SNR) of the image by increasing the obtained photon flux while maintaining the original spatial resolution of STED. In addition, the SNR and resolution of prSTED are strongly correlated with the intensity of depletion beam. Corresponding theoretical and experimental analysis about this feature are also conducted. In general, considering the enhanced signal strength, imaging speed and compatibility with some other imaging techniques, we believe prSTED would be a helpful promotion in biomedical imaging.**

Fluorescence microscopy plays an important role in biomedical imaging and analysis due to its specificity and non-invasive advantages. Among this, confocal laser scanning (CLSM) microscopy has been widely used benefiting from its excellent optical sectioning ability and better resolution than conventional diffraction-limited microscopy (theoretically up to $\sqrt{2}$ times) [1]. On this basis, in order to further improve the resolving ability of microscopic system, a variety of super-resolution microscopies have emerged in the past few decades [2-8]. Among these techniques, STED can effectively improving the spatial resolution through forcing the point spread function (PSF) of confocal microscopy to be compressed and narrowed by introducing another saturated hollow depletion beam in the excitation light path of confocal [3]. In current biomedical applications, the actual resolution can generally reach 20-70 nm, approximately three to ten times higher than confocal [9]. However, since STED achieves resolution improvement by suppressing the effective fluorescence of the periphery of confocal PSF through saturated hollow depletion beam, a considerable part of the original effective fluorescence is quenched by the depletion beam, and the signal intensity in actual imaging is also susceptible. Another method is image scanning microscopy (ISM), which has recently been frequently studied [10-15]. Unlike STED, ISM changes the detection method of confocal and can effectively enhances the signal strength as well as achieving a resolution increase of up to √2 times relative to confocal [8]. ISM uses wide-field detector or point detector array, together with photon reassignment algorithm (somewhere also called pixel reassignment) to collect the fluorescence signal and process original images [12-14]. Consequently, the final ISM image is of enhanced SNR and resolution, which effectively solves the problem of resolution and SNR constraints caused by the introduction of pinhole in conventional confocal. Here, with avalanche diode (APD) detector array to detect the fluorescence signal, we implement STED based on array detection and photon reassignment. We briefly explain the imaging principle and characteristics of prSTED. Simulations and experimental results show that prSTED can effectively improve the imaging signal strength while maintaining the original high resolution of STED. Further, a notable problem in prSTED is that the intensity of depletion light has a great impact on the system PSF which ultimately affects the imaging characteristics of prSTED. We also implement a detailed investigation into this issue. In addition, the imaging speed can also get improved benefitting from enhanced signal strength.

prSTED is mainly based on the traditional STED method, but the difference is that the array detection method substitute for the single-point detection in traditional STED. Thus, each detector in detector array will have its own sub-PSF and corresponding subimage. Due to the different spatial locations of each detectors, each sub-PSF has a certain degree of deformation and positional shift

$$sPSF_i = PSF_{exc} \times (PSF_{emi} \otimes PIN_i) \quad (1)$$

where $PSF_{exc}$ and $PSF_{emi}$ represent the excitation and emission PSF of the imaging system respectively, $PIN_i$ and $sPSF_i$ represent the pinhole and sub-PSF of *i*th detector. On this basis, through the photon reassignment algorithm, each sub-STED image is shifted according to their respective shift vector, and then all the subimages are added together to obtain the final prSTED image

$$I_{prSTED} = \sum_{i=1}^{n} I_i(\boldsymbol{r} - \boldsymbol{r}_i) \quad (2)$$

where $r_i$ and $I_i$ represent the shift vector and sub-STED image of ith detector, $I_{prSTED}$ is the final prSTED image. Ideally, assuming that the imaging system aberration and the Stokes shift between fluorescence excitation and emission is ignored, shift vector is completely determined by the spatial position of each detector. Further, recent researches have shown that shift vector can be determined by calculating the correlation degree between each subimage and the central detector subimage [15]. Therefore, the reconstructed image can be recovered only through all subimages, and there is no need to know the system parameters and Stokes shift information in advance. Also, it has good compatibility with aberrations in actual imaging. In the following simulations and experiments, we would always use this correlation calculation method.

In actual STED imaging, as the commonly used detector, APD have a certain operating bandwidth limitation that each time a detector receives a photon, it will not be able to receive other photons for a certain period of time. This results in the system not being able to detect all signal photons and reducing the maximum photon flux that can be reached. On the other hand, considering the quantum properties of photon, the position of each photon on the detection surface would be random, and the signal photons will be distributed in a Poisson distribution. Thus, for array detection, even if some detectors cannot detect the photons arriving at a certain moment, the photons can still be detected normally by another detector, thereby effectively improving the actual photon count rate and signal strength. Specifically, within one laser pulse period T, the detected signal intensity for one APD is;

$$I_{single} = I_0 \times k = I_0 \times \left(1 - e^{-\mu}/\mu\right) \quad (3)$$

where $I_0$ and $I_{single}$ represent the ideal and practical signal intensity of single detector respectively. k is the attenuation coefficient introduced by detector's working bandwidth limitation, and ideally k is taken as 1 [16]. $\mu$ is the fluorescent photon rate, which represent the counts of emitted photons in one period T. In the case of array detection, considering the quantum properties of photon, the total signal intensity detected by all detectors is

$$I_{array} = \sum_{i=1}^{n} p_i I_i = \sum_{i=1}^{n} \frac{I_0(1 - e^{-\mu p_i})}{\mu} \quad (4)$$

where $p_i$ is the probability that the $i$th detector detects a certain photon. Here, the actually used detector array consists of 19 separate APDs with a spatial distribution shown in Fig. 1a. The pinhole size for each APD is 0.2 AU, and the total size of 19 pinholes roughly equals to a circular pinhole with the diameter of 1 AU. The photon detection efficiency as a function of photon rate is shown in Fig. 1b, where the red and blue curves represents array detectors and single detector respectively. The result indicates that as the emitted fluorescence increases, the difference between photon detection ability of two methods becomes more and more obvious, where the detection efficiency of array detectors is 1.2 to 3.8 times higher than single detector as $\mu$ changes from 0.5 to 5. As for STED imaging, Fig. 1c shows the respective sub-PSF of 19 detectors. Due to the difference in the position of the pinhole of each detector, the shape and center position of each sub-PSF is slightly different compared to the center sub-PSF. If adding all the sub-PSFs directly, the system PSF of STED based on array detection (adSTED) can be obtained as shown in Fig. 1e2, and its peak signal intensity is 1.5 times higher than that of STED with single detector (Fig. 1e1). For simplicity, here we assume that $\mu$ is a constant in the whole image and takes a value of 1, which means that one photon is emitted in each laser pulse period (same processing is adopted in whole simulation of this paper). Besides, the saturation depletion intensity in STED simulation model introduced here is 11.4mW [17], and the depletion intensity used here is 100mW.

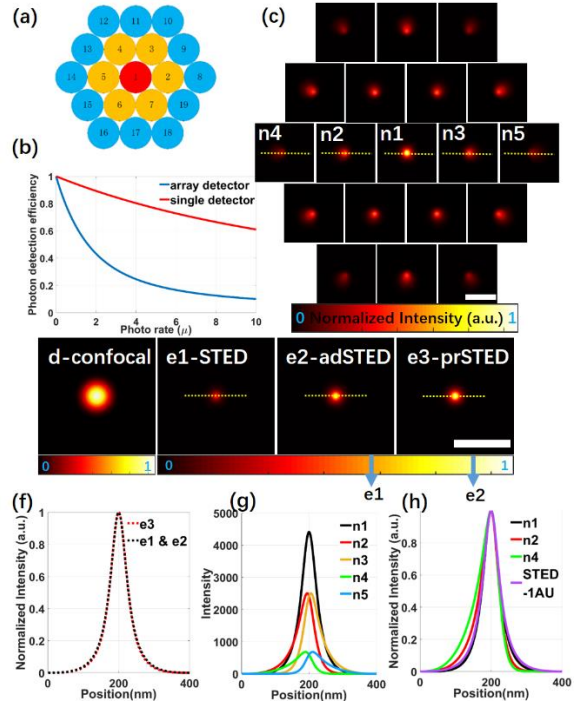

Fig. 1. Simulation of array detection and prSTED. (a) Spatial distribution of 19 detectors. (b) Photon detection efficiency as a function of photon rate. (c) Sub-PSFs of 19 detectors. For visual clarity, 12 outside sub-PSFs are doubled in intensity. Scale bar: 250 nm. (d) and (e) PSF of confocal, STED, adSTED and prSTED, where the last three are normalized to the maximum of prSTED and the peak intensity of STED and adSTED is marked in the colorbar with blue arrows. Scale bar: 500 nm. (f) Line profile along the center of PSF of STED, adSTED and prSTED, marked with yellow dashed line in (e1)-(e3). (g) Line profile along the center of the PSFs of five detectors (n1-n5, labelled in (c) with character and yellow dashed line). (h) Line profile along the center of the PSFs of three detectors (n1, n2 and n4), normalized to respective maximum and shifted according to shift vector. STED of 1AU is introduced for reference.

Further, using photon reassignment algorithm to process the images of all detectors, the system PSF of prSTED can be obtained, and the PSFs of STED, adSTED and prSTED are shown side by side in Fig. 1e. Here all three are normalized to the maximum of prSTED and corresponding peak intensity are marked by blue arrows in the colorbar. It is clearly that prSTED can further improve the signal intensity on the basis of adSTED with 1.1 times peak intensity. On the other hand, Fig. 1f depicts the full width at half maximum (FWHM) of the PSF of STED, adSTED and prSTED, and all three are almost the same size, both 56nm. This is because the FWHM of PSF of STED is not sensitive to the position and size of the pinhole, so it does not change much after reassignment. This can be seen from Fig. 1g and 1h, where profiles of the sub-PSFs of five detectors marked with yellow dashed line in Fig. 1c are depicted (profiles here are normalized to the maximum value of center sub-PSF and their respective maximums in Fig. 1g and 1h respectively, and in Fig. 1h they are shifted according to shift vectors and PSFs of STED at 1 AU are also introduced for reference). Meanwhile, since the peak positions of shifted PSFs overlap together, it can further increase the signal intensity of adSTED. Consequently, prSTED can enhance imaging SNR compared to traditional STED and their resolution keep almost the same. This is very important for actual STED imaging. Since STED in principle deliberately discards many original fluorescent signal in exchange for improved resolution, prSTED method can just compensate for this inherent defect. Moreover, since image post-processing algorithms, such as deconvolution, are usually sensitive to noise, the enhanced SNR can also help to further improve the imaging resolution during image processing.

The schematic of our prSTED system is depicted in Fig. 2. The main body of the system is based on a traditional STED system. Two continuous lasers of 488 nm (MCLS1, Thorlabs) and 592 nm (VFL-P-1000-592-OEM1, MPB Communications Inc.) are used as light sources in the system for excitation and depletion light. Two beams after spatial co-alignment are finally focused by the 1.4NA objective lens (OL) and then illuminate the sample. Fluorescence emitted by the sample is collected by the same OL and finally detected by the detection fibers and array detectors.

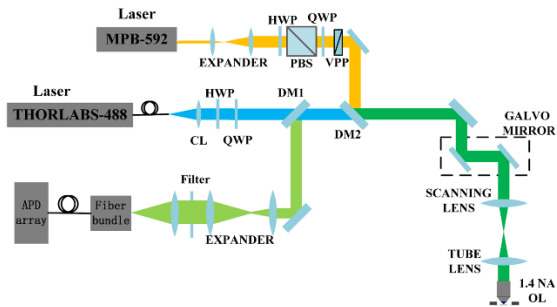

Fig. 2. Setup of the prSTED system. The fluorescence is collected through the fiber bundle and detector array.

Unlike traditional STED system, a detector array consisting of 19 APDs with a 20MHz working bandwidth replaces the traditional single-point detector, while a bundle of 19 fibers is used to collect fluorescence emitted by the sample and transfer to their respective APD. Here, each fiber is equivalent to a separate pinhole with a core diameter of 0.2 AU, and 19 fibers arranged in a hexagonal manner form a circular pinhole of approximately 1 AU in size (shown in Fig. 1a). In addition, before fluorescence is focused to the fiber end, we add a five-fold expander system to match the size of Airy disk at the end of the fiber bundle to the core diameter of the fiber bundle.

In experiment, we first tested the imaging performance of prSTED with 100 nm fluorescent particles. Confocal and STED images were measured with a single APD and a single fiber with a large core diameter of 1 AU, and prSTED image with the fiber bundle of 19 fibers and corresponding 19 APDs shown in Fig. 1a. As shown in Fig. 3, the resolution of STED and prSTED are quite close, both significantly higher than that of confocal with a depletion intensity of 200mW, and Fig. 3e shows the FWHM of 10 random particles with a statistical mean value of 121.1 nm, 123.3 nm and 253.9 nm for prSTED, STED and confocal respectively. However, the photon detection efficiency of prSTED and STED show significant differences. As shown in Fig. 3a and 3b, the photon counts per 20μs of prSTED is 1178, approximately 3 times higher than that of STED, indicating that the signal strength is effectively enhanced in prSTED. This is much clearer in the enlarged windows marked through white boxes in Fig. 3a and 3b, where the two enlarged parts are normalized to the common maximum of two images. Besides, the resolution can be further improved through Richard-Lucy deconvolution and the statistical mean FWHM is 110 nm as shown by the red histogram in Fig. 3e.

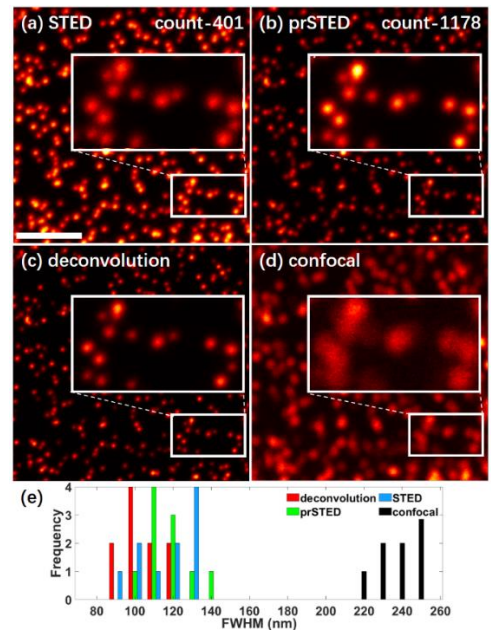

Fig. 3 Experimental results of prSTED through imaging 100 nm fluorescence particles. (a)-(d) STED, prSTED, RL deconvolution and confocal images. The insets are enlarged views of images in the white boxes, and the enlarged views in a and b are normalized to the common maximum of the two images. The dwell time is 20 μs and pixel size is 15 nm. (e) Histograms of the FWHM of ten random particles. Corresponding results are shown in different colors for confocal, STED, prSTED and deconvolution.

Next, we further analyze the effect of depletion intensity on the performance of prSTED. This effect can be mainly divided into two aspects: (i) the emitted fluorescence intensity (or the photon rate $\mu$) would decrease with increasing depletion intensity, which weakening the detection efficiency gain obtained through array detector; (ii) the signal strength improvement through photon reassignment algorithm would also be weakened under higher depletion intensity. We conduct a detailed investigation into this effect through both simulations and experiments, and for simplification, we suppose that photon rate $\mu$ still keeps a constant of 2 in simulations. Fig. 4a shows the signal gain (prSTED versus STED) and resolutions as a function of depletion intensity. It can be seen that, at lower depletion intensity, prSTED can improve the resolution to some extent, but the resolution of two methods will quickly tend to be the same with increasing depletion intensity. In terms of signal strength, prSTED can enhance the signal strength of the image, but this gain will also decay with the increase of depletion intensity and eventually tend to be a certain value of 2.1 (which is the gain of array detection relative to single point detection). That is to say, if ignoring array detection and only considering photon reassignment algorithm, the signal strength gain will be significantly restricted by the depletion intensity. This is because as the intensity increases, the system PSF is continuously compressed and narrowed, which decreases the sensitivity of sub-PSFs upon the position of pinhole. Hence, the peak position of each sub-PSF continuously gathers toward the center, and consequently leading to a weakened signal gain after photon reassignment. This point can be clearly seen from Fig. 4b where each point in the figure indicates the peak position of each sub-PSF, and for contrast, we connect the points under the same depletion intensity with a line of certain color. As intensity increases, the peaks positions continue to move closer and closer to the center, and at the end almost all gather to the center position.

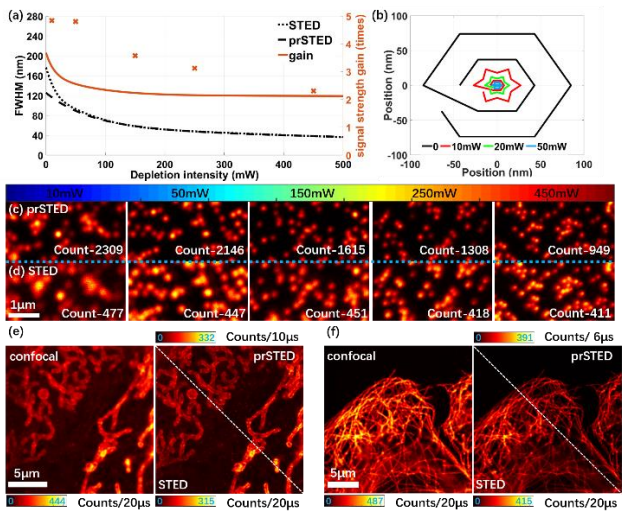

Fig. 4. Simulation and experiment analysis of the relationship between depletion intensity and prSTED imaging SNR and resolution. (a) Simulation profiles of depletion intensity versus signal strength gain, STED resolution and prSTED resolution. Left Y-axis is resolution and the right is signal gain. The five discrete data points are experimental data for signal gain of prSTED. (b) Positions of peak intensities of each sub-PSF under different depletion intensity, which represent the shift vectors corresponding of each detector. For better contrast, we connect each point together with a string and each color corresponds to a certain depletion intensity. (c)-(d) STED and corresponding prSTED images of 100 nm fluorescence particles under five depletion intensities. (e)-(f) Confocal, STED and prSTED images of mitochondria and microtube, respectively. Corresponding photon count and dwell time are indicated above and below the images.

We verified above analysis through imaging 100 nm particles at several different depletion intensity of 10, 50, 150, 250, and 450 mW. As shown in Fig. 4c and 4d, under 10mW depletion intensity, the photon count of prSTED is 2309, indicating a 4.8 times peak intensity enhancement over STED. However, as depletion intensity increases, the intensity gain gradually decreases and is 2.3 with 450mW depletion intensity, which is very close to the simulation result. Corresponding experimental data is marked in Fig. 4a through discrete points. The experimental results are generally consistent with simulation analysis, and the differences lies in that the actual photon rate decreases as increasing depletion intensity but is always higher than that in simulation, especially under lower depletion intensity.

Moreover, benefitting from enhanced photon counts and signal strength, the imaging speed can also get improved. Here we utilized our system to image Oregon Green 488 labelled mitochondria and Alexa plus 488 labelled microtube under depletion intensity of 100 mW. Fig. 4e and 4f shows corresponding confocal, STED and prSTED images side by side. For mitochondria in Fig. 4e, the photon counts for confocal and STED within 20 μs is 444 and 315 respectively; while for prSTED, it can obtain almost the same photon counts within 10 μs, and the imaging SNR and quality also keep very well. As for microtube, due to its stronger fluorescence intensity, it can achieve similar signal intensity only within 6 μs. Hence, the imaging speed is improved for 2 to 3.3 times while maintaining image quality and resolution.

In conclusion, we propose a novel STED method based on array detection and photon reassignment. Theoretical analysis and experimental results show that this method can improve the imaging SNR while maintaining the spatial resolution. Meanwhile, this is beneficial for further processing of images by post-processing methods, such as subtractive imaging and various deconvolution algorithms [18, 19]. Since these algorithms are generally sensitive to imaging noise levels, the enhanced signal strength can be further converted to resolution improvement. Also, the enhanced photon flux can transform to improved imaging speed and up to 3.3 times improvement is realized. Besides, simulations and experimental results indicate that depletion intensity have a significant impact on imaging characteristics of prSTED and actual performance would get restricted under intense depletion intensity. In addition, here we implement prSTED based on CW configuration, but this method is also feasible for pulsed STED, and the signal enhancement would be more significant for low repetition rate system such as T-Rex STED [20, 21]. This is because emitted photons of this scheme are much more concentrated within one prolonged laser pulse period, and consequently generating a larger photon rate μ. Furthermore, since APD is capable of acquiring the time information of detected photons, our system can also be applied to other related imaging methods, such as time-gated detection, fluorescence lifetime imaging (FLIM), and fluorescence correlation spectroscopy [22, 23]. Specially, since time-gated detection rejects part of the photons through the introduction of time gate, the feature of prSTED to increase the photon flux can just make up the signal loss. For FLIM imaging, array detection can alleviate the pile-up effect of single detector, thus reducing the limitation of low illumination intensity and long photon collection time, and effectively improving the imaging speed of FLIM.

**Funding.** National Natural Science Foundation of China (NSFC) (61827825, 61427818，and 61735017); National Basic Research Program of China (973 Program) (2015CB352003); Natural Science Foundation of Zhejiang province (LR16F050001); and Fundamental Research Funds for the Central Universities (2018FZA5005).